\documentclass[twocolumn,prl,showpacs,color,superscriptaddress,epsfig]{revtex4}
\usepackage[french]{babel}
\usepackage[latin1] {inputenc}
\usepackage{epsfig}
\usepackage{array}
\usepackage{amsmath}

\begin{document}

\title{Antiferromagnetic interactions in  single crystalline Zn$_{1-x}$Co$_{x}$O thin films}
\author{P. Sati}

\affiliation{Laboratoire Mat{\'e}riaux et Micro{\'e}lectronique de
Provence,   13397, Marseille Cedex 20, France}

\author{C. Deparis}
\affiliation{ Centre de Recherche sur l'Hétéro-Epitaxie et ses
Applications-CNRS, 06560, Valbonne Sophia-Antipolis, France }

\author{C. Morhain}
\affiliation{ Centre de Recherche sur l'Hétéro-Epitaxie et ses
Applications-CNRS, 06560, Valbonne Sophia-Antipolis, France }

\author{S. Sch\"afer}
\affiliation{Laboratoire Mat{\'e}riaux et Micro{\'e}lectronique de
Provence,   13397, Marseille Cedex 20, France}
\author{A. Stepanov}
\affiliation{Laboratoire Mat{\'e}riaux et Micro{\'e}lectronique de
Provence,   13397, Marseille Cedex 20, France}

%****************************************************************************************************************************

\begin{abstract}

In a rather contradictory situation regarding magnetic data on
Co-doped ZnO, we have succeeded in fabricating high-quality single
crystalline Zn$_{1-x}$Co$_{x}$O ($x=0.003-0.07$) thin films. This
gives us the possibility, for the first time, to examine the {\it
intrinsic} magnetic properties of ZnO:Co at a quantitative level
and therefore to address several unsolved problems, the major one
being the nature of the Co-Co interaction in the ZnO structure.
\end{abstract}
\pacs{ 75.30.Et, 75.30.Cr, 75.70.-i}

\maketitle

%*********************************************************************

Manipulating  the spin of an electron rather than its charge opens
fascinating new routes for information storage and processing. A
quantum computer operating on electron-spin qubits is probably one
of the  most appealing challenges in the new field of research and
applications known as spintronics \cite{nielsen}. With the use of
inherently quantum-mechanical effects, much faster computation can
be achieved,  as compared to its classical counterpart. Another
particularly striking example of this emerging spintronics
technology is the giant magnetoresistive read head  for hard disk
drive, which gave rise to a  hundred-fold increase in hard disk
capacity during the last decade \cite{ohno}. All devices of this
kind are currently made of multilayer {\it metallic} hetero- or
tunnelling structures; however, the successful realization of the
spin manipulation in {\it semiconductor} structures would open up
the way to numerous totally new fields of applications including
quantum information processing \cite{zutic}. This is why diluted
magnetic semiconductors (DMS) have become a focus of considerable
interest in recent years.

DMS of the II-VI group,  have attracted
much attention as essential materials for practical semiconductor
spintronic devices such as spin filters \cite{ohno} or spin
polarizers \cite{ohno1}. In particular, the theoretical
predictions  based on the
local spin density approximation (LSDA), triggered extensive
studies of ZnO:TM alloys with a special focus  on ZnO:Co as the
most promising candidate for a room-temperature ferromagnetic (FM)
semiconductor \cite{sato}. Many experiments have been reported on
this material fabricated by a variety of methods
\cite{song,venkatesan,neal,tuan,kittilstved,
kane,yin,alaria,bouloudenine,kolesnik,kim,han,risbud,jedrecy};
however, the magnetic properties of ZnO:Co  still remain a
controversial issue since the observed magnetic behavior appears
to be  strongly dependent on the preparation methods and is poorly
reproducible.

 Ferromagnetism was reported for thin films
and bulk samples of ZnO:Co with a very large spread of spontaneous
magnetic moment from 6.1 $\mu_B$/Co to 0.01 $\mu_B$/Co accompanied
by a Curie temperature well above room temperature
\cite{song,venkatesan,neal,tuan,kittilstved, kane}. At the same
time and with the use of practically the same preparation methods,
the absence of ferromagnetism and paramagnetic behavior down to
helium temperatures in ZnO:Co were claimed by many authors
\cite{yin,alaria,bouloudenine,kolesnik,kim,han,risbud,jedrecy}.
 Interestingly, these last results seem to be less
process-dependent and  most of them indicate dominant {\it
antiferromagnetic} (AFM) interactions between Co ions  (negative
sign of $\theta$ in the Curie-Weiss law) while a positive sign of
$\theta(x)$ was also observed at low Co concentration
\cite{risbud,jedrecy}. It is worth noting that numerous secondary
phases, such as metallic Co \cite{kane,kolesnik}, CoO \cite
{kolesnik}, Co$_3$O$_4$ \cite{risbud} and ZnCo$_2$O$_4$
\cite{kolesnik}, were found to occur in this material, which
further complicates the interpretation of experimental data. %[Note
%also that the presence of secondary phases in the form of small
%(nanosize) clusters makes them very difficult to be detected by
%many conventional structural-analysis methods.]

On the theoretical side, the situation with ZnO:Co is not better.
There exists a certain consensus that LSDA  % , in its simplest form,
has difficulties when applied to the magnetic state of TM-doped
ZnO, since it does not account for correlations between
d-electrons and leads almost "automatically" to a semimetallic FM
ground state. Quite surprisingly, an improved version of LSDA,
LSDA+$U$, which is supposed to be free from these deficiencies,
leads to controversial results as regards the exchange constant
sign between   Co$^{2+}$ ions in ZnO. Indeed, in their recent
paper, Chanier {\it et al.} \cite{chanier} show that the exchange
constants, $J^{out}$ and $J^{in}$, between nearest-neighbor (NN)
Co ions in the ZnO wurtzite structure are both negative (AFM) and
have the values  $-$9 K and $-$21 K, respectively. In contrast,
Lee and Chang \cite{lee} and Sluiter {\it et al.} \cite{sluiter}
have detected a competition between FM $J^{out}$ and AFM $J^{in}$
in Co-doped ZnO. Quite close to this result are those of Risbud
{\it et al.} who found that the FM and AFM ground state in ZnO:Co
have almost the same energy. %This fact, in their opinion, could
%explain the experimentally observed change of $\theta(x)$ at low
%$x$, mentioned above \cite{risbud}.

 In this Letter, we report the magnetic measurements we performed on
single crystalline Zn$_{1-x}$Co$_{x}$O thin films. The use of
single crystalline films allows us to circumvent many problems
inherent to polycrystalline samples (secondary phases,
imperfections, surface and structural defects, etc.) and, for the
first time, to perform a thorough analysis of the {\it intrinsic}
magnetic properties of ZnO:Co at a quantitative level, thereby
addressing most of the questions raised above. We first
demonstrate the absence of ferromagnetism in these compounds.
Next,  by comparing the magnetization measurements with our
cluster model we conclude decisively about the NN exchange
constants, which we find both antiferromagnetic in this material.
We also show that  these conclusions hold even under strong
n-doping.

 Zn$_{1-x}$Co$_{x}$O thin films ($x$ varying from
 0.003 up to 0.07)
were grown on sapphire substrates by plasma-assisted MBE and had
thicknesses of about 1$\mu$m, the $c$-axis of the wurtzite
structure being perpendicular to the film plane. The conductivity
of the films was $n$-type, with residual carrier concentrations
$n_{e}$ $<$ 10$^{18}$ cm$^{-3}$.  For some of them, the electron
doping level ($n_{e}$$\geq$10$^{20}$cm$^{-3}$) was controlled by
in-situ gallium doping  (Zn$_{1-x}$Co$_{x}$O:Ga). Two-dimensional
growth  was achieved for a growth temperature of 560°C (i.e., 50°C
higher than the optimal growth temperature used for ZnO),
resulting in streaky reflection high-energy electron diffraction
patterns and very smooth surfaces (rms $\leq$ 1nm). The full
widths at half maximum of the x-ray rocking curves measured in
high-resolution $\omega$ scan position, were in the range of
$\omega$ $\sim$ 0.15° along (002), (105), and (-105). The low,
identical values of $\omega$ measured both for (105) and (-105),
indicated a large column diameter, close to $\omega$. The full
width at half maximum along (101), measured in skew position, was
$\sim$ 0.75° corresponding to a twist value of $\pm$ 0.5°. While
the column diameters remained large, $\omega$ values were found to
increase slightly and gradually with the Co concentration.

Magnetic measurements were performed using a Quantum Design MPMS
XL magnetometer in magnetic fields up to 50 kOe and in the
temperature range $2-300$ K. Typically, the observed magnetic
moment was relatively small, 10$^{-3}$$-$10$^{-4}$ emu and,
therefore, special attention was paid to the subtraction of
spurious contributions. A pure ZnO film on a sapphire substrate
and a sample holder were examined separately and their signals
were subsequently subtracted from the total magnetic moment. The
Co content $x$ of the studied samples was determined by energy
dispersive x-ray (EDX) microanalysis and was found to be quite
uniform as the average dispersion $\Delta x$ was around 0.003, for
a large number of scanned "spots" (N$\simeq$ 30). For the lowest
concentration ($x<0.005$), the $x$ value was determined by
magnetic measurements. Additionally, low-temperature EPR in the
X-band was used to check that Co$^{2+}$ ions occupy tetrahedral
substitutional sites in the wurtzite structure and to ensure the
absence of secondary magnetic phases \cite{L2MP}.

%%%%%%%%%%%%%%%%%%%%%%%%%%%%%%%FIG1%%%%%%%%%%%%%%%%%%%%%%%

\begin{figure}[t]

      \includegraphics[bb=22 15 283 215 width=.3,height=.25\textheight]{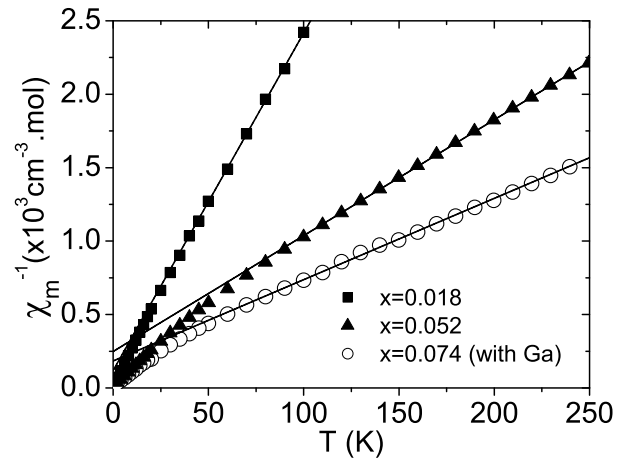}
      \caption{ Temperature dependence of the inverse magnetic susceptibility
      for Zn$_{x}$Co$_{1-x}$O samples with $x=0.018$,
      $x=0.052$ and $x=0.074$ taken at $H=10$ kOe and $H\perp c$.}
      \label{MdeT}

\end{figure}
In order to facilitate the ensuing discussion, we first present
our theoretical model. The $^4$A$_2$ ground state of an isolated
Co$^{2+}$ at a tetrahedral site of the ZnO host lattice is
described by an $S=3/2$ spin Hamiltonian determined by only three
constants \cite{jedrecy, estle, L2MP}: the two $g$-factors,
$g_{\parallel}=2.236$ and $g_{\perp }=2.277$, and the zero-field
splitting constant $D=2.76$ cm$^{-1}$, which are used  in
this work as fixed parameters.

As the Co concentration is increased, the role played by Co-Co
interactions becomes more important and a model which considers an
ensemble of isolated Co$^{2+}$ fails. The most straightforward way
to account for these interactions is provided by the so called
nearest-neighbor cluster model \cite{behringer,shapira,gratens}.
%Since this model is widely employed in DMS physics, here we
%briefly recall only its main features.
In this model only the
largest NN exchange constants are included and all other exchange
constants are set equal to zero. As was demonstrated with ZnO:Mn
\cite{gratens}, two groups of NN's have to be distinguished in the
ZnO lattice: six "\emph{in plane}" NN's which are in the same $c$
plane as the central cation (the corresponding exchange constant
labelled $J^{in}$) and six "\emph{out of plane}" NN's which are
coupled to the central cation by $J^{out}$.

Here, four cluster types are considered (singles, pairs, open
triangles and closed triangles). This allows a quite accurate
calculation of the magnetic moment for Co concentrations up to
0.05 (where 90\% of all  Co$^{2+}$ spins are counted).

The magnetization of a cluster of type $\alpha$ is obtained from
the usual equation: $  M_{\alpha}(H,T)= (\partial
F_{\alpha}/\partial H)_T$ where $F_{\alpha}$ is the free energy of
the cluster found by exact  numerical diagonalisation of its
Hamiltonian, which contains  a Heisenberg exchange interaction
term, a single-ion anisotropy, and the Zeeman energy. The field
and temperature dependent magnetization thus becomes $
M(H,T)=\sum_{\alpha}P_{\alpha}(x)M_{\alpha}(H,T)/N_{\alpha} $,
where $P_{\alpha}(x)$ is the probability \cite{behringer} that a
Co ion belongs to a  cluster of type $\alpha$ , and $N_{\alpha}$
is the number of Co ions in this cluster.

We now turn to the results of our magnetic measurements. In
Fig.\ref{MdeT} we show the temperature dependence of the inverse
static magnetic susceptibility, $\chi^{-1}(T)$, measured at $H=10$
kOe and $H \perp c$  for three samples. The film for $x=0.074$ was
co-doped with Ga in order to reach the electron concentration
$n_{e}\approx 10^{20}$cm$^{-3}$, which  was further confirmed by
electrical measurements. We found, however, that the n-doping does
not substantially affect the magnetic properties of ZnO:Co films.
A linear increase of $\chi^{-1}(T)$ at higher temperatures was
fitted to the Curie-Weiss law $\chi^{-1}(T)=(T-\theta(x))/C(x)$.
As a typical example, $\theta=-60\pm30 K$ for $x=0.074$ can be
cited. An unusually large error bar arises mainly from the
uncertainty in the background contribution because the ratio of
the ZnO:Co signal to the background one is about  $1/20$ at high
temperatures. The situation becomes even worse at lower $x$
rendering a proper determination of $\theta$ impossible.
%The situation becomes even worse at lower $x$: while
%the mean value of $\theta$ is found to be negative, its standard
%deviation is too large to allow us to conclude definitely about
%its sign.
Therefore, we reach the conclusion that the accuracy of
our measurements at high temperature is not sufficient to perform
a quantitative analysis of the exchange interactions in
Zn$_{1-x}$Co$_x$O films. Moreover, it is quite clear that the
value of $\theta$ cannot help much in obtaining NN exchange
constants, since in Zn$_{1-x}$Co$_x$O it also contains
contributions from the single-ion anisotropy. As is shown below,
the low-temperature magnetic measurements combined with the
cluster model turn out to be  much more informative from this
point of view.
%%%%%%%%%%%%%%%%%%%%%%%%%%%%%%FIG2%%%%%%%%%%%%%%%%%%%%%%%%%%%%%%%%%%%
\begin{figure}[hb]
\includegraphics[bb=18 15 285 220 width=.3,height=.25\textheight]{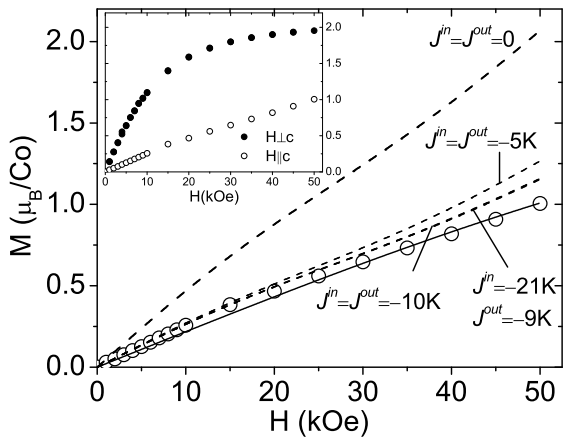}
\caption{$M$ vs $H$ plot for a Zn$_{1-x}$Co$_x$O sample with
$x=0.052$ at $T=2$ K  and $H\parallel c$ (open circles). Dashed
lines represent the calculated magnetization for
$J^{out}=J^{in}=$0, $-$5 K, $-$10 K and $J^{out}=-9$ K,
$J^{in}=-21$ K. The solid line is computed using $J^{out}=-9$ K,
$J^{in}=-21$ K and $T_0=3$ K. The inset shows $M(H)$ for both
orientations, $H\parallel c$ and $H\perp c$.}\label{MdeH}
\end{figure}

In Fig. \ref{MdeH} we show the magnetization of the
Zn$_{1-x}$Co$_x$O film with $x = 0.052$ as a function of magnetic
field at $T = 2$ K. As expected from the results on weakly
Co-doped ZnO films \cite{L2MP}, the $M(H)$ curves reveal a
considerable magnetic anisotropy of Co$^{2+}$ in the wurtzite
lattice (see the inset in Fig. \ref{MdeH}). This could explain, at
least in part, the difficulties in the interpretation of magnetic
experiments on polycrytalline ZnO:Co samples when one uses a
simple Brillouin function in order to fit $M(H)$ curves
\cite{han,yin,risbud}.

In order to probe the exchange interactions in ZnO:Co, we have
compared the experimental data with the results of simulations
performed using the following values for the NN exchange
constants: $J^{in} = J^{out} = 0$, $-5$ K, $-10$ K and $J^{in}
=-21$ K and $J^{out} =-9$ K, all other parameters being fixed. As
seen from Fig. \ref{MdeH}, a curve which corresponds to an
ensemble of isolated Co$^{2+}$ ions ($J^{in}=J^{out}=0$) passes
significantly higher  as compared to the experimental data. A much
better agreement is observed when one uses a negative and
increasing value of $J$. The calculated $M(H)$ does not depend on
$J$ if both $|J^{in}|$ and $|J^{out}|$ exceed $10$ K and remain
negative. However, the closest curve to the experimental data
still passes above the measured points.

Now we would like to improve our cluster model by taking into
account weak exchange interactions between single Co$^{2+}$ spins,
which means that the effect of the distant-neighbor exchange
interactions in ZnO lattice must be considered in some way or
another. To do this we shall use %a popular concept in the DMS
%field, namely,
the effective-temperature approximation which
corresponds physically to replacing the Curie susceptibility of
noninteracting singles by the Curie-Weiss one, i.e. to replacing
the actual temperature $T$ by a higher effective temperature
$T_{eff}$ = $T + T_0$ \cite{gaj,shapira}. Including this parameter
in the model, a better match with the experimental data is
obtained. The calculated curve for $T_0$ = 3 K and $J^{in} =-21$
K, $J^{out} =-9$ K is shown by the solid line in Fig. \ref{MdeH}.
Note that at low fields ($10<H< 20$ kOe) there exists a small but
reproducible "excess" of the experimental magnetization as
compared with the calculated curve. We attribute this to
particular magnetic states arising from weak interactions between
distant neighbors.

\begin{figure}[th]
    \begin{center}
      \includegraphics[bb=18 15 280 220 width=.3,height=.25\textheight]{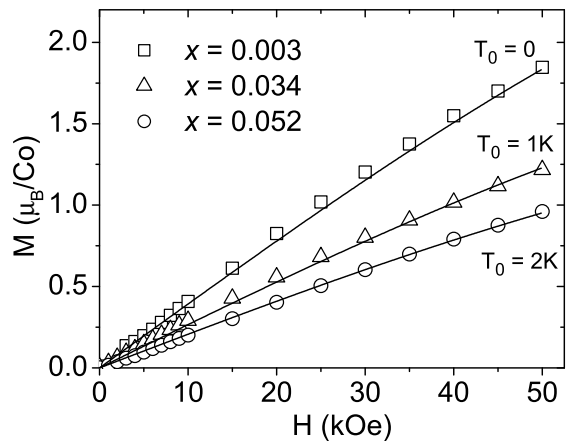}
      \caption{$M$ vs $H$ plots for Zn$_{1-x}$Co$_x$O films with $x$ =
      0.003 (squares), $x$ = 0.034
      (triangles) and $x$ = 0.052
      (circles) at $T$ = 4 K and $H \parallel c$. Solid lines are calculations
       using $J^{in} =-21$ K and $J^{out} =-9$ K
      with $T_0$ as an  adjustable parameter.
       }\label{MdeHcomp}.
  \end{center}
\end{figure}

Further evidence in favor of the existence of AF Co-Co
interactions in ZnO is provided by the $M(H)$ dependence on $x$.
Figure \ref{MdeHcomp} displays the experimental data for samples
with $x$ = 0.003, $x$ = 0.034 and $x$ = 0.052 taken at $T$ = 4 K
and $H\parallel c$, as well as the simulated curves. As above, the
exchange constants were kept fixed at $J^{in} =-21$ K, $J^{out}
=-9$ K, and $T_0$ was adjusted to obtain the best agreement. The
slope of the $M(H)$ curves is seen to decrease continuously with
increasing $x$, indicating that the magnetization per Co site
decreases as the Co concentration is increased. This fall-off in
magnetization was earlier observed in
polycrystalline \cite{yin,risbud,kim,bouloudenine} %and epilayer \cite{kimbis}
ZnO:Co, where it was attributed to the formation of
antiferromagnetic clusters of Co$^{2+}$ ions. Note the increase of
$T_0$ with the Co concentration, which is quite in line with the
mean-field approximation \cite{denissen,gaj}.

\begin{figure}[h]
    \begin{center}
      \includegraphics[bb=18 18 283 218 width=.3,height=.25\textheight]{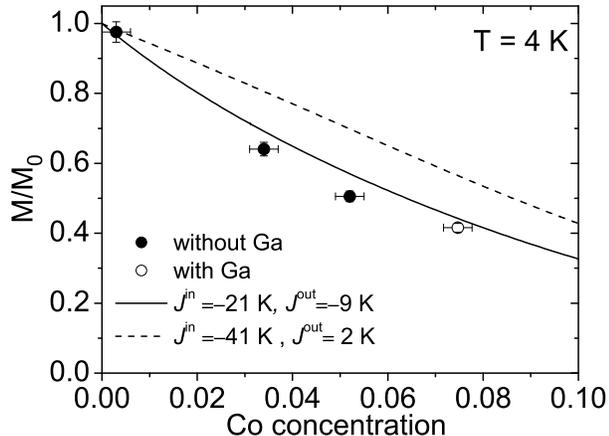}
      \caption{Concentration dependence of magnetization measured at $T = 4$
      K, $H \parallel c$
      and $H = 50$ kOe for four Zn$_{1-x}$Co$_{x}$O films with and
      without co-doping with Ga. Solid and dashed lines  correspond to
      the two scenarios discussed in the text.
       }\label{ratio}.
  \end{center}
\end{figure}

We now return to the question regarding the sign of the exchange
constants in ZnO:Co, namely, which of the theoretically suggested
scenarios, 1)$J^{in}=-21$ K and $J^{out}=-9$ K according to
Ref.\onlinecite{chanier}, or 2)$J^{in}=-41$ K and $J^{out}=2$ K
according to Ref.\onlinecite{lee}, is realized in Co-doped ZnO? In
fact we have already partly confirmed the former scenario, when
analyzing the data on Fig.2. Nevertheless, in order to render our
analysis more convincing, we have plotted in Fig. \ref{ratio} the
observed magnetization at $H =50$ kOe and $T=4$ K divided by
$M_{0}$, the calculated magnetization at $H =50$ kOe,  $T=4$ K and
for $J^{in}=J^{out}=0$, as a function of the Co concentration. The
two scenarios are presented in Fig. \ref{ratio}, where the solid
line  corresponds to  the first one and the dashed line represents
the second \cite{rem1}. Obviously, this is the case where both
$J's$ are negative, which is realized in ZnO:Co. It is  worth
noting, once again, that, as follows from  Fig. 4 and the above
discussion, our ZnO:Co films which were co-doped with Ga  do not
present any substantial difference in their magnetic behavior as
compared with the Ga-free samples.

%**************************************************************************************************************************
%\section{Conclusions}
In summary, in a rather controversial situation regarding the
magnetic data on Co-doped ZnO published so far in the literature,
we have succeeded in preparing high-quality single crystalline
Zn$_{1-x}$Co$_{x}$O ($x=0.003-0.07$) thin films grown by
plasma-assisted molecular beam epitaxy. We find that ZnO:Co is
paramagnetic down to helium temperatures over the studied range of
$x$. We focus then on the magnetization process at low
temperature.  In order to describe the reduction of the observed
magnetization with increasing $x$, we have developed a
phenomenological cluster model%, parameterized by the NN exchange
%constants,
which is favorably compared with our experimental data.
This has enabled us, by analyzing the concentration dependence of
the ZnO:Co magnetization, to safely conclude about the NN exchange
constants, $J^{in}$ and $J^{out}$, which are found to be both
antiferromagnetic and to exceed 10 K in this material. We also
show that this conclusion holds even under strong n-doping.

Clearly, the most interesting problem, which now remains open, is
reproducible p-doping and the way it affects the magnetic
properties of Co-doped ZnO.

%%%%%%%%%%%%%%%%%%%%%%%%%%%%%%%%%%%%%%%%%%%%%%%%%%%%%%%%%%%%%%%%%%%%%%%

\end{document}